\documentclass[desactivate]{aa}
\usepackage[varg]{txfonts}
\usepackage{graphicx}
\graphicspath{{images/}} 
\usepackage[version=3]{mhchem} 
\usepackage{pifont} 
\usepackage{longtable}
\bibpunct{(}{)}{;}{a}{}{,} 
\usepackage{color}
\AddToHook{begindocument/before}{\RequirePackage{hyperref}}

\begin{document}
	
	\title{The factors that influence protostellar multiplicity} 
	\subtitle{II. Gas temperature and mass in Perseus with APEX}
	\author{N. M. Murillo\inst{1,2},  C. M. Fuchs\inst{3}, D. Harsono\inst{4}, T-.H. Hsieh\inst{5,6}, D. Johnstone\inst{7,8}, R. Mignon-Risse\inst{9} \and M. V. Persson\inst{10}, N. Sakai\inst{2}}
	
	\institute{Instituto de Astronomía, Universidad Nacional Autónoma de México, AP106, Ensenada CP 22830, B. C., México\\ \email{nmurillo@astro.unam.mx}
		\and Star and Planet Formation Laboratory, RIKEN Pioneering Research Institute, Wako, Saitama 351-0198, Japan
		\and Fox2Space - FTSCO, The Fault Tolerant Satellite Computer Organization, Weigunystrasse 4, 4040 Linz, Austria
		\and Institute of Astronomy, Department of Physics, National Tsing Hua University, Hsinchu, Taiwan
		\and Taiwan Astronomical Research Alliance (TARA), Taiwan
		\and Academia Sinica Institute of Astronomy and Astrophysics, No. 1, Section 4, Roosevelt Road, Taipei 10617, Taiwan
		\and NRC Herzberg Astronomy and Astrophysics, 5071 West Saanich Rd, Victoria, BC, V9E 2E7, Canada
		\and Department of Physics and Astronomy, University of Victoria, Victoria, BC, V8P 5C2, Canada
		\and Department of Physics, Norwegian University of Science and Technology, NO-7491 Trondheim, Norway
		\and Independent researcher, Sagas Väg 5, 43431 Kungsbacka, Sweden}
	
	\abstract
	{Protostellar multiplicity is a common outcome of the star formation process. To fully understand the formation and evolution of these systems, the physical parameters of the molecular gas together with the dust must be systematically characterized.}
	{Using observations of molecular gas tracers, we characterize the physical properties of cloud cores in the Perseus molecular cloud (average distance of 295 pc) at envelope scales (5000 to 8000 AU).}
	{We used Atacama Pathfinder EXperiment (APEX) and Nobeyama 45m Radio Observatory (NRO) observations of \ce{DCO+}, \ce{H2CO} and \ce{c-C3H2} in several transitions to derive the physical parameters of the gas toward 31 protostellar systems in Perseus. The angular resolutions ranged from 18$\arcsec$ to 28.7$\arcsec$, equivalent to 5000 -- 8000 AU scales at the distance of each subregion in Perseus. Gas kinetic temperature was obtained from \ce{DCO+}, \ce{H2CO} and \ce{c-C3H2} line ratios. Column densities and gas masses were then calculated for each species and transition. Gas kinetic temperature and gas masses were compared with bolometric luminosity, envelope dust mass, and multiplicity to search for statistically significant correlations.}
	{Gas kinetic temperature derived from \ce{H2CO}, \ce{DCO+} and \ce{c-C3H2} line ratios have average values of 26 K, 14 and 16 K, respectively, with a range of 10 -- 26 K for \ce{DCO+} and \ce{c-C3H2}. The gas kinetic temperature obtained from \ce{H2CO} line ratios have a range of 13 -- 82 K. Column densities of all three molecular species are on the order of 10$^{11}$ to 10$^{14}$ cm$^\mathrm{-2}$, resulting in gas masses of 10$^{-11}$ to 10$^{-9}$ M$_{\odot}$. Statistical analysis of the physical parameters finds: i) similar envelope gas and dust masses for single and binary protostellar systems; ii) multiple ($>$2 components) protostellar systems tend to have slightly higher gas and dust masses than binaries and single protostars; iii) a continuous distribution of gas and dust masses is observed regardless of separation between components in protostellar systems.}
	{}

	
	\titlerunning{Perseus gas mass and temperature vs. multiplicity}
	\authorrunning{Murillo et al.}
	
	\maketitle

	\section{Introduction}
	\label{sec:intro}
	
	Most stars are found to be in multiple stellar systems, with companion frequencies increasing with mass (see \citealt{offner2022ppvii} and references therein). This demonstrates that multiplicity is a central aspect and general outcome of the star formation process across time.
	Observations of starless cores in various star forming regions have sought to identify substructures linked to fragmentation (Perseus: \citealt{schnee2010}; Taurus: \citealt{hacar2011,caselli2019}; Ophiuchus: \citealt{kirk2017}; Chamaeleon I: \citealt{dunham2016}). These studies have mixed results, with the lack of substructure attributed to evolutionary stage, the mechanism of collapse, or observational limitations.
	Dust continuum studies have provided information on protostellar multiplicity and companion fractions, distribution of components within cloud cores, and speculated on formation mechanisms (e.g., \citealt{looney2000,chen2013,mairs2016,tobin2016,sadavoy2017,encalada2021,tobin2022,offner2022ppvii}).
	While these studies provide a rough temperature and mass estimate from the dust, they lack detailed information on the physical conditions of the gas.
	Various key parameters of star formation that are essential ingredients for simulations and linking observations to theoretical studies can only be provided by atomic and molecular gas observations (e.g., \citealt{mairs2014,luo2022,chen2024,neralwar2024}).
	
	The Perseus molecular cloud has been extensively studied. 
	A few examples of studies from the last decade have explored cloud structure (e.g., \citealt{hacar2017,chen2020}); line emission (e.g., \citealt{higuchi2018,hsieh2019,yang2021,tafalla2021,dame2023}); multiplicity (e.g., \citealt{pineda2015,tobin2016,murillo2016,sadavoy2017,pokhrel2018}); and polarization (e.g., \citealt{doi2021,coude2019,choi2024}).
	Given this wealth of information, Perseus is an excellent region to study the factors that affect multiplicity in star formation.
	Furthermore, the Perseus molecular cloud presents a range of star forming environments. 
	Externally irradiated subregions like IC348 and NGC1333, contrast with isolated environments like L1455, B1 and L1448. Clustered subregions, like NGC1333, and non-clustered environments, like L1448, are also found in the Perseus molecular cloud \citep{plunkett2013}.
	The distance to these subregions varies between 279 and 302 pc, with an average distance of 294 pc \citep{zucker2018}. 
	More recent work using the LAMOST DR8, 2MASS PSC and Gaia EDR3 data \citep{cao2023} find consistent distances for Perseus.
	
	A pilot study of ten protostellar systems in Perseus with the Atacama Pathfinder EXperiment (APEX, \citealt{gusten2006}) was carried out with the aim of studying the temperature-multiplicity relation at envelope scales ($\sim$8000 AU, \citealt{murillo2018PaperI}, hereafter Pilot Paper).
	The observations targeted \ce{DCO+}, \ce{H2CO}, \ce{c-C3H2} and \ce{^{13}CO} in the 230 and 360 GHz APEX bands.
	This study suggested a lack of correlation between temperature and multiplicity, but instead suggested a mass-multiplicity relation.
	However, statistically significant conclusions could not be obtained due to the sample size of ten protostellar systems.
	
	Expanding the sample size to 37 protostellar systems observed with on-the-fly (OTF) maps using the Nobeyama 45m Radio Observatory (NRO) and targeting \ce{HNC}, \ce{HCN}, \ce{N2H+} and \ce{HCO+} enabled a more statistically significant study of the temperature and mass relations with respect to protostellar multiplicity at scales of 5000 AU (\citealt{paperII2024}, hereafter Paper I). 
	Gas kinetic temperature was derived using the $I$(\ce{HCN})/$I$(\ce{HNC}) $J$=1--0 ratio \citep{graninger2014,hacar2020}, and compared to \ce{NH3} temperatures \citep{friesen2017} and dust temperatures \citep{zari2016}.
	The study was further complemented with APEX \ce{HNC} $J$=4--3 single-pointing observations in order to derive \ce{H2} volume density from the \ce{HNC} $J$=4--3 / $J$=1--0 ratio.
	No relation between multiplicity and gas kinetic or dust temperature was found in Paper I, as suggested in the Pilot Paper.
	A relation between mass and degree of multiplicity was found in Paper I, with a continuous mass distribution independent of component separation.
	This finding suggests a continuum in formation mechanisms for multiple protostellar systems, rather than distinct core and disk mechanisms for close and wide separations in multiple protostellar systems.
	This result is consistent with the existence of equal mass wide binaries (e.g., \citealt{elbadry2019}), outspiraling due to gas accretion causing binaries to widen (e.g., \citealt{mignonrisse2023}), and dynamical capture with efficient inspiraling (e.g., \citealt{kuruwita2023}).

	The aim of the present study is to explore the relation between multiplicity, mass and temperature using molecular species that trace the inner regions and structures of protostellar cores.
	Different molecular species probe different structures and temperature regimes within protostellar cores (e.g., \citealt{tychoniec2021}).
	In protostellar cloud cores where \ce{CO} is abundant, formylium (\ce{DCO+}) is enhanced when \ce{CO} begins to freeze-out onto dust grains.
	In protostellar envelopes, \ce{DCO+} is found between the \ce{CO} freeze-out temperature (20 to 30 K, \citealt{jorgensen2005}) and the desorption density of \ce{CO} ($\sim$10$^{4}$ cm$^{-3}$).
	It is useful to note that \ce{DCO+} also has a warm formation path at gas temperatures above 30 K, mostly present around the disk for embedded protostars \citep{favre2015}.
	Previous observations showed that warm \ce{DCO+} is more readily detected in the $J$=5--4 transition with interferometric data, but is less prominent in single dish observations \citep{murillo2018}.
	This means that the \ce{DCO+} emission in the present study traces the cool regions of the protostellar envelope (e.g., \citealt{mathews2013,murillo2015}).
	Formaldehyde (\ce{H2CO}) is an excellent thermal probe of gas between 30 and 100 K, with gas kinetic temperature increasing proportional to the line ratio. 
	Irradiated environments (e.g., outflow cavities) are best traced with species such as cyclopropenylidene (\ce{c-C3H2}; e.g., \citealt{drozdovskaya2015,murillo2018PaperI,tychoniec2021}). 
	Each set of molecular line emission ratios will provide an independent measurement of gas kinetic temperature averaged over the beam area of the observations.
	\ce{DCO+} $J$=5--4 / $J$=3--2 will provide the gas kinetic temperature of the cold inner envelope. 
	The \ce{H2CO} and \ce{c-C3H2} ratios can provide gas kinetic temperatures of the warm inner envelope and irradiated outflow cavity walls, respectively.
	The derived temperatures will be compared to source parameters (bolometric luminosity, dust mass, and multiplicity) to search for correlations.
	Because each species traces a different structure within the cloud core, they can be used to probe the amount and distribution of material available for accretion around protostellar sources (envelope versus outflow cavity).
	The per-species beam-averaged column density and gas mass derived from \ce{DCO+}, \ce{H2CO} and \ce{c-C3H2} thus enables comparison of masses, per species, versus multiplicity in the inner regions of protostellar cores.

	This work presents single-dish observations toward five subregions in the Perseus molecular cloud.
	The observations target 31 protostellar systems using APEX single-pointing observations of \ce{DCO+}, \ce{H2CO} and \ce{c-C3H2}, complemented with NRO OTF maps of \ce{DCO+} $J$=1--0.
	Observational data details are provided in Sect.~\ref{sec:observations}. 
	Section~\ref{sec:results} describes the results of the observations. 
	Derivation of physical parameters and statistical analysis is explained in detail in Sect.~\ref{sec:analysis}. 
	Finally, the discussion and conclusions of this work are provided in Sects.~\ref{sec:discussion} and \ref{sec:conclusions}.

    \begin{table*} 
    	\centering
    	\caption{Molecular species in this work}
    	\begin{tabular}{c c c c c c c c c}
    		\hline \hline
    		Molecule & Transition & Frequency & E$_{\rm up}$ & log$_{10}$ A$_{\rm ij}$ & Beam & Peak range & RMS noise range & Line width range \\
    		& & GHz & K & & $\arcsec$ & K & K & km~s$^{-1}$ \\
    		\hline
    		\multicolumn{9}{c}{OTF Map: Nobeyama} \\
    		\hline
    		\ce{DCO+} & 1--0 &  72.03931 &  3.46 & -4.66 & 23 & 0.6 -- 5 & 0.1 -- 0.5 & 0.4 -- 1 \\
    		\hline
    		\multicolumn{9}{c}{Single pointing: APEX} \\
    		\hline
    		\ce{DCO+} & 3--2 & 216.11258 & 20.74 & -2.62 & 28.7 & 0.3 -- 3.3 & 0.01 -- 0.07 & 0.4 -- 1 \\
    		\ce{c-C3H2} & 3$_{3,0}$--2$_{2,1}$ & 216.27876 & 19.47 & -3.33 & 28.7 & 0.05 -- 0.8 & 0.01 -- 0.07 & 0.5 -- 2 \\
    		\ce{c-C3H2}\tablefootmark{a} & 6--5 & 217.82215 & 38.61 & -3.23 & 28.7 & 0.05 -- 0.8 & 0.01 -- 0.07 & 0.5 -- 2 \\
    		\ce{c-C3H2} & 5$_{1,4}$--4$_{2,3}$ & 217.94005 & 35.42 & -3.35 & 28.7 & 0.05 -- 0.8 & 0.01 -- 0.07 & 0.5 -- 2 \\
    		\ce{p-H2CO} & 3$_{0,3}$--2$_{0,2}$ & 218.22219 & 20.96 & -3.55 & 28.7 & 0.09 -- 2 & 0.01 -- 0.07 & 0.4 -- 4 \\
    		\ce{p-H2CO} & 3$_{2,2}$--2$_{2,1}$ & 218.47563 & 68.09 & -3.80 & 28.7 & 0.07 -- 0.2 & 0.01 -- 0.07 & 0.4 -- 4 \\
    		\ce{p-H2CO} & 3$_{2,1}$--2$_{2,0}$ & 218.76007 & 68.11 & -3.80 & 28.7 & 0.04 -- 0.2 & 0.01 -- 0.07 & 0.4 -- 4 \\
    		\ce{DCO+} & 5--4 & 360.16978 & 51.86 & -2.42 & 18 & 0.1 -- 2 & 0.02 -- 0.15 & 0.4 -- 1 \\
    		\ce{p-H2CO} & 5$_{0,5}$--4$_{0,4}$ & 362.73602 & 52.31 & -2.86 & 18 & 0.1 -- 1 & 0.02 -- 0.15 & 0.4 -- 4 \\
    		\hline
    	\end{tabular}
    	\\
    	\tablefoot{\tablefoottext{a}{Contains both ortho- and para forms.}}
    	\tablebib{All rest frequencies were taken from the Cologne Database for Molecular Spectroscopy (CDMS; \citealt{CDMS_2016}).
    		The \ce{DCO+} entries are based on \citet{DCO+_rot_2005}. The \ce{c-C3H2} 
    		entry was based on \citet{c-C3H2_isos_rot_1987} with transition frequencies 
    		important for our survey from \citet{c-C3H2_rot_1986} and from \citet{c-C3H2_isos_rot_2012}. 
    		The \ce{H2CO} entries are based on experimental data from \citet{h2co1996}.}
    	\label{tab:lines}
    \end{table*}

    \begin{table*} 
    \centering
    \caption{Classification schemes for statistical analysis}
    \begin{tabular}{c c p{0.3\linewidth} p{0.3\linewidth}}
    	\hline \hline
    	Scheme & Bins & Selection criteria & Formation mechanism considered \\
    	\hline
    	1 & Full sample & All observed pointings & None \\
    	2 & Singles, binaries, multiples & Binaries have separations of $<$7$\arcsec$ (2100 AU at $d~=$~300 pc) & Core vs disk fragmentation \\
    	3 & Singles, binaries, multiples & Binaries have 2 components regardless of separation & None \\
    	4 & Singles, multiples & Multiples have $\geq$2 components regardless of separation & None \\
    	\hline
    \end{tabular}
    \\
    \label{tab:schemes}
    \end{table*}

	\section{Observations}
	\label{sec:observations}
	
	\subsection{Sample}
	\label{subsec:sample}
	The source sample included in this work is composed of 31 protostellar systems, selected based on the region overlap between APEX and the NRO OTF maps of \ce{DCO+} $J$=1--0.
	The sample contains the ten protostellar systems from the Pilot Paper (hereafter, pilot sample) to complement newer APEX observations (hereafter, extended sample).
	The mini cluster NGC1333 IRAS4, observed in the NRO OTF map but not in the APEX single pointings, is included in the sample. 
	The sample in this work contains 19 multiple protostellar systems and 12 single protostellar sources distributed in five subregions of Perseus: NGC1333, L1448, L1455, B1, and IC348.
	Among the 19 multiple protostellar systems, eight systems are close binaries, five systems are wide binaries, and six systems are higher order multiples (three or more components within a common cloud core).
	Close and wide binaries are defined following the criteria used in \citet{murillo2016}, namely that close binaries are those with separations $<$7$\arcsec$ (2100 AU at $d$ = 300 pc) and are not resolved in \textit{Herschel} PACS \citep{poglitsch2010} observations.
	A pointing with APEX was made for each component of a protostellar system that can be resolved in \textit{Herschel} PACS images.
	This implies that higher-order multiples and wide binary systems have several pointings per system, while close binaries and single protostars have one pointing per system. Higher-order multiples can have unresolved components, for example, NGC1333 SVS13A and L1448 N-B have components with separations $<$7$\arcsec$.
	Hence, the 31 systems included in the sample were observed with 46 pointings.
	Table~\ref{tab:source} lists the systems and pointings used in this work.
	
	\subsection{Dust continuum parameters: $L_{bol}$ \& $M_{env}$}
	\label{subsec:dustcontparam}
	Two parameters obtained from dust continuum observations are used in this work: bolometric luminosity $L_{bol}$ and envelope dust mass $M_{env}$. 
	The $L_{\rm bol}$ parameter is obtained from the spectral energy distribution (SED) of the source using trapezoidal integration with the equation 
	\begin{equation}
		L_{\rm bol} = 4~\pi~d^2~\int_{0}^{\nu'}~S_{\nu}~d\nu
	\end{equation}
	where $d$ is the distance and $S_{\nu}$ is the flux at a given frequency $\nu$. The SEDs are taken from \citet{murillo2016}, which include Spitzer, \textit{Herschel} PACS and (sub)millimeter fluxes.
	For any two components that have separations $\geq$7$\arcsec$, their $L_{bol}$ values are for the individual protostars. In contrast, components with separations $<$7$\arcsec$ have $L_{bol}$ values which are the arithmetic average of the luminosity of both protostellar components. The 7$\arcsec$ limit is due to the spatial resolution of \textit{Herschel} PACS data used in the SEDs. A detailed exploration of bolometric luminosity toward unresolved protostellar component pairs is given in \citet{murillo2016}.
	
	For the envelope dust mass we use the method from \citet{jorgensen2004} which is given by
	\begin{equation}
		\label{eq:dustmass}
		M_{\rm env} = 0.44 M_{\odot}~\left(\frac{L_{\rm bol}}{1~L_{\odot}}\right)^{-0.36}~\left(\frac{S_{850\mu m}}{1~{\rm Jy~beam}^{-1}}\right)^{1.2}~\left(\frac{d}{125~{\rm pc}}\right)^{1.2}
	\end{equation}
	where $M_{\rm env}$ is calculated from the 850 $\mu$m dust continuum peak intensity $S_{\rm 850\mu m}$.
	The James Clerk Maxwell Telescope (JCMT) Gould Belt Survey SCUBA-2 data release 3 (GBS DR3\footnote{The data was downloaded through \url{https://doi.org/10.11570/18.0005}}; \citealt{kirk2018,chen2016,pattle2025}) 850$\mu$m maps were used to obtain the peak fluxes for the sources in our sample. 
	Since previous papers (Pilot paper and Paper I) used peak fluxes from the COMPLETE survey \citep{ridge2006}, a comparison of the envelope dust masses derived from SCUBA-2 and SCUBA peak fluxes is presented in Appendix~\ref{app:contdata}.
	
	For both parameters, $L_{bol}$ and $M_{env}$, the adopted distances per subregion are 288 pc for L1448, 299 pc for both NGC1333 and L1455, 301 pc for B1, and 295 pc for IC348 \citep{zucker2018}.
	Both $L_{\rm bol}$ and $M_{\rm env}$ are listed in Table~\ref{tab:source} and included in the CDS online data for the current work.

	\subsection{Atacama Pathfinder EXperiment (APEX)}
	\label{subsec:APEXobs}
	
	The pilot sample of ten protostellar systems was observed with the Swedish Heterodyne Facility Instrument (SHeFI; \citealt{nystrom2009,vassilev2008shefi}) receivers APEX-1 (O-098.F-9320B.2016, NL GTO time) and APEX-2 (M-099.F-9516C-2017) on 1 December 2016 and 7 -- 12 July 2017, respectively.
	The extended sample was observed using the PI230 and First Light APEX Submillimeter Heterodyne (FLASH$^{\rm +}$; \citealt{klein2014flash}) receivers in single pointing mode.
	Observations with PI230 (O-0104.F-9307A) were carried out on three dates: 8 July, 11 July and 12 October 2019.
	FLASH$^{\rm +}$ observations (O-0104.F-9307B) were done on six dates: 11, 15--18, 20 October 2019.
	The extended sample targeted 27 protostellar systems, but only 21 are included in this work due to overlap with the NRO OTF maps of \ce{DCO+}.
	
	The receivers APEX-1 and APEX-2 were used to observe the pilot sample in two spectral settings with central frequencies 217.11258 and 361.16978 GHz, and a bandwidth of 4 GHz. Full details are given in the Pilot Paper and briefly described here.
	The APEX-1 spectral setup targeted \ce{DCO+} $J$=3--2, three transitions of \ce{H2CO}, and three transitions of \ce{c-C3H2}.
	APEX-2 was used to observe \ce{DCO+} $J$=5--4 and \ce{H2CO} $J$=5--4.
	Typical noise levels range between 15 to 70 mK for APEX-1, and 20 to 100 mK for APEX-2, for a channel width of 0.4 km s$^{-1}$, and HPBW of 28.7$\arcsec$ and 18$\arcsec$, respectively.
	Adopted beam efficiencies were $\eta_{\rm mb}$ = 0.75 and 0.73, for APEX-1 and APEX-2, respectively.
	
	The central frequency of PI230 was set to 217.11258 GHz, mainly targeting \ce{DCO+} $J$=3--2, three transitions of \ce{H2CO}, and three transitions of \ce{c-C3H2}. 
	The 7.5 GHz bandwidth of PI230 also included additional molecular species (\ce{DCN} $J$=3--2, \ce{C^{18}O} $J$=2--1, \ce{^{13}CO} $J$=2--1, two transitions of \ce{SO}, and one transition of \ce{CH3OH}) which will be treated in future work.
	Typical noise levels for the observations ranged between 10 to 30 mK for a channel width of 0.4 km s$^{-1}$ and a HPBW of 28.7$\arcsec$. 
	The beam efficiency $\eta_{\rm mb}$ for PI230 observations at 230 GHz is 0.8.
	
	The spectral setup with FLASH$^{\rm +}$ targeted \ce{DCO+} $J$=5--4, \ce{H2CO} $J$=5--4, \ce{DCN} $J$=5--4, and \ce{HNC} $J$=4--3, with a central frequency of 361.16978 GHz and a bandwidth of 4 GHz.
	The \ce{HNC} $J$=4--3 data has been reported in Paper I, and \ce{DCN} $J$=5--4 will be discussed in a future work.
	Typical noise levels for the observations ranged between 18 and 145 mK for a channel width of 0.4 km s$^{-1}$, and a HPBW of 18$\arcsec$.
	A beam efficiency of $\eta_{\rm mb}$ = 0.73 was adopted, based on APEX-2 beam efficiency at 352 GHz.
	
	The GILDAS/CLASS software package\footnote{\url{https://www.iram.fr/IRAMFR/GILDAS/}} \citep{gildas2013ascl} was used to perform standard calibration.
	All spectra are averaged to 0.4 km~s$^{-1}$ in order to increase the sensitivity. 
	Sample spectra for three systems are briefly discussed in Appendix~\ref{app:spectra} and shown in Fig.~\ref{fig:APEX_spec}.
	The averaged spectra results in detection in at least two channels for the weakest lines, and three or more channels for stronger line emission.
	Spectral line parameters (peak, width, rms) for all APEX data were measured with the GILDAS/CLASS gaussian fitting routines.
	Targeted molecular species and their respective parameters are listed in Table~\ref{tab:lines}.
	
	The system Per25, an isolated single protostellar source in L1455, was included in both the pilot and extended samples (Fig.~\ref{fig:Per25_spec}).
	Hence it is used to compare the quality of both datasets. 
	The extended sample has lower noise (20 -- 60 mK with $\Delta v$ =  0.4 km~s$^{-1}$) relative to the pilot sample (25 -- 100 mK with $\Delta v$ = 0.4 km~s$^{-1}$).
	In the extended sample observations, six molecular lines are detected toward L1455 Per25 with $>3\sigma$: \ce{DCO+} $J$=3--2, \ce{HNC} $J$=4--3, \ce{H2CO} $J$=3$_{0,3}$--2$_{0,2}$, and the three transitions of \ce{c-C3H2}.
	In the pilot observations, only two detections were above 3$\sigma$ for L1455 Per25, \ce{DCO+} $J$=3--2 and \ce{HNC} $J$=4--3, while the others were undetected or marginal ($<3\sigma$).
	Thus the extended sample observations provide a better constraint for single protostars than the pilot sample observations.
	
	\subsection{Nobeyama 45m Radio Observatory (NRO)}
	\label{subsec:NROobs}
	
	The \ce{DCO+} $J$=1--0 observations of Perseus with the NRO were made using the FOur-beam REceiver System on the 45m Telescope (FOREST; \citealt{minamidani2016}) frontend and the Spectral Analysis Machine for the 45 m Telescope (SAM45;\citealt{kamazaki2012}) backend.
	Observations (G22019 PI: N. M. Murillo) were carried out between February and April 2023. 
	The objective of these observations was to make OTF maps similar to those presented in Paper I but targeting the corresponding deuterated species. 
	In the current work, only the \ce{DCO+} $J$=1--0 spectral data averaged over 23$\arcsec$ beams is used. The remaining data will be presented in a future work.
	
	For the \ce{DCO+} OTF map, the channel resolution is 30.52 kHz ($\sim$0.1 km~s$^{-1}$) with a bandwidth of 125 MHz.
	The OTF map pixel size is 6$\arcsec$.
	Noise levels ranged between 0.1 and 0.5 K, with an angular resolution of 23$\arcsec$. 
	We adopted a beam efficiency of $\eta_{\rm mb}$ of 0.6 based on the data provided in the NRO website.
	The value was interpolated from the measurements by NRO for the 2022--2023 observing season\footnote{\url{https://www.nro.nao.ac.jp/~nro45mrt/html/prop/eff/eff2023.html}}, since the frequency of \ce{DCO+} $J$=1--0 (72 GHz) is not standard operation for FOREST.
	
	Standard data calibration and imaging for each datacube were done with the NOSTAR software package provided by NRO. Gaussian fitting to extract spectral line peak, width and noise level from the map were performed using the bettermoments Python library \citep{teague2018,teague2019}. 
	
	\begin{figure*} 
		\centering
		\includegraphics[width=0.97\linewidth]{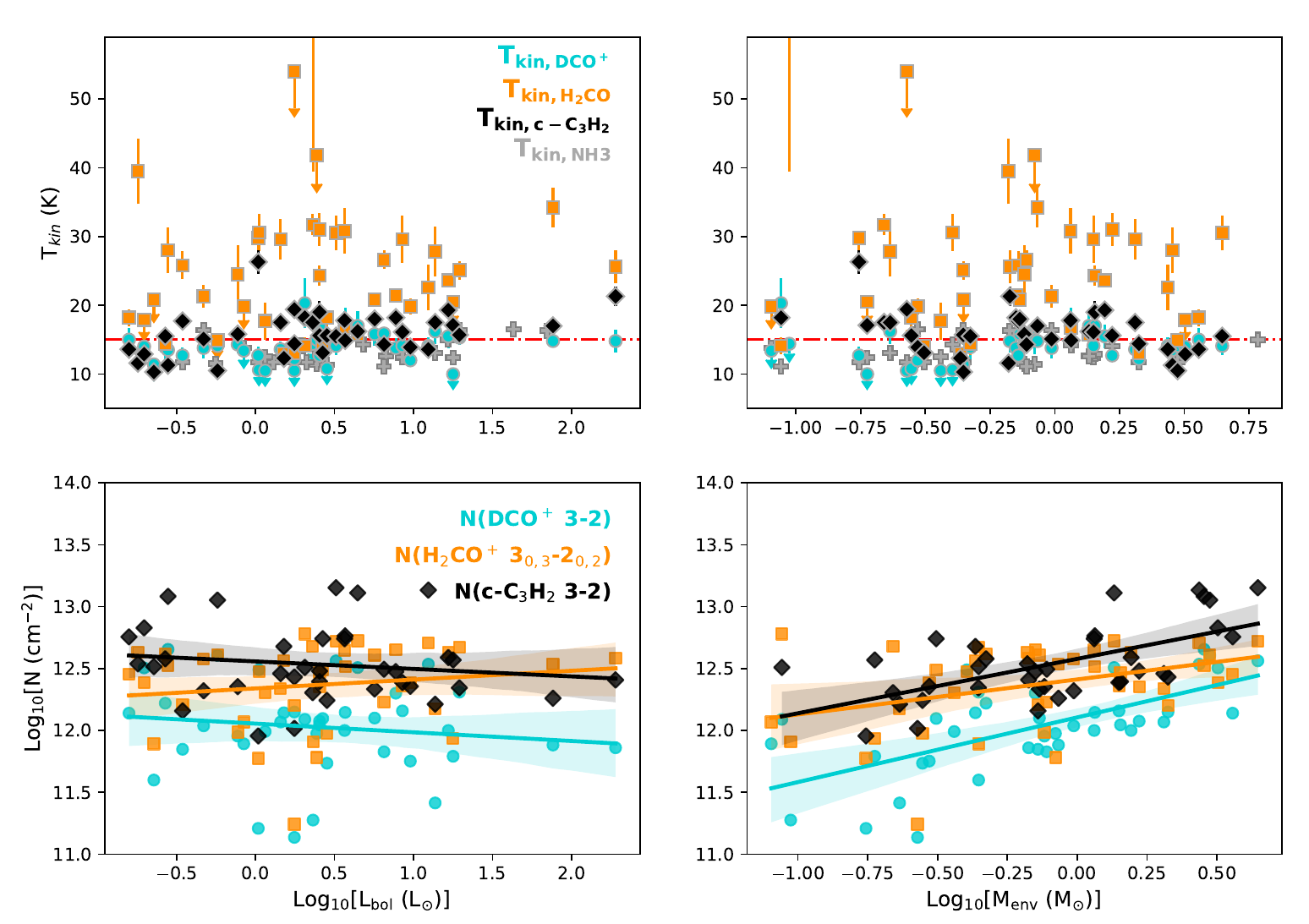}
		\caption{Physical parameters derived from APEX observations. \textit{Top row}: Gas kinetic temperature and corresponding errors obtained from molecular line ratios. Cyan circles show the temperature obtained from \ce{DCO+} $J$=5--4 / $J$=3--2, orange squares show the average temperature obtained from the three \ce{H2CO} ratios, and the black diamonds show the average temperature obtained from the two \ce{c-C3H2} ratios. Upper limits are indicated with a downward arrow. The horizontal red dash-dotted line indicates the average gas kinetic temperature ($T_{kin}~\sim$15 K) derived from the I(\ce{HCN})/I(\ce{HNC}) $J$=1--0 (Paper I). The gray crosses show the gas kinetic temperature from \ce{NH3} observations ($T_{kin, \ce{NH3}}$) obtained from the core catalog of \citet{rosolowsky2008}. Left and right panels plot gas kinetic temperature versus bolometric luminosity ($L_{\rm bol}$) and envelope dust mass ($M_{\rm env}$), respectively.
			\textit{Bottom row}: Column density for \ce{DCO+} (cyan circles), \ce{H2CO} (orange squares) and \ce{c-C3H2} (black diamonds) obtained from the respective lowest transition in the APEX data. Solid lines and shaded areas show the linear regression for the data with the corresponding color. Left and right panels plot column density versus $L_{\rm bol}$ and $M_{\rm env}$. Error bars are smaller than the symbols.
		}
		\label{fig:TN_LM}
	\end{figure*}
	
	\begin{figure*} 
		\centering
		\includegraphics[width=0.97\linewidth]{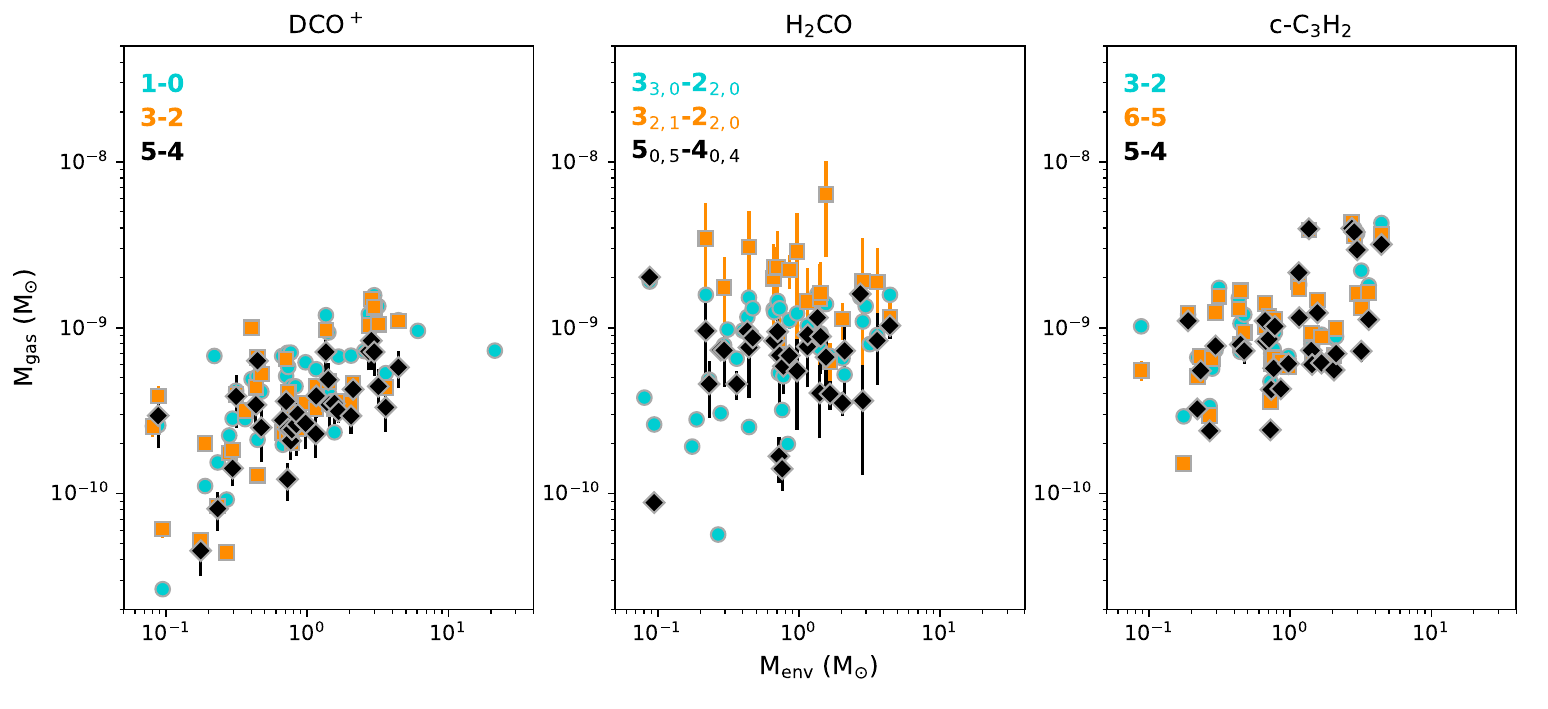}
		\caption{Gas mass ($M_{\rm gas}$) versus dust mass ($M_{\rm env}$) for all transitions of \ce{DCO+} (\textit{left panel}), \ce{H2CO} (\textit{center panel}), and \ce{c-C3H2} (\textit{right panel}). This plot shows that the gas masses from \ce{DCO+} and \ce{c-C3H2} are consistent between transitions, while \ce{H2CO} shows a broader spread. In the center panel, the gas mass from \ce{H2CO} 3$_{2,2}$--2$_{2,1}$ is not shown to avoid a crowded plot, but follows a similar distribution as the other transitions shown in the panel.}
		\label{fig:Mass_fullSample}
	\end{figure*}
	
	\begin{figure*} 
		\centering
		\includegraphics[width=0.974\linewidth]{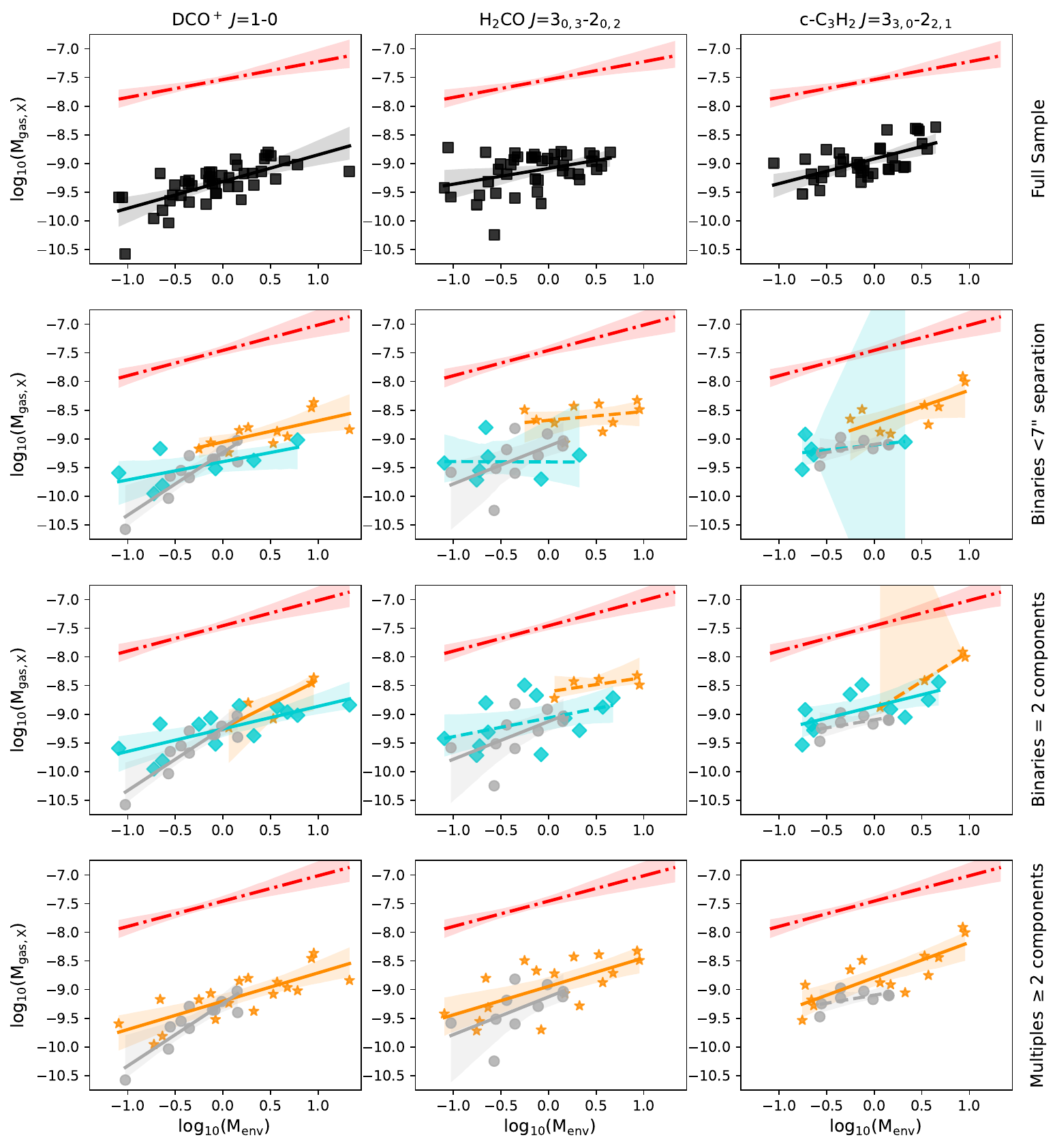}
		\caption{Relations of the \ce{DCO+} $J$=1--0 gas mass ($M_{\rm \ce{DCO+}}$, \textit{left column}), \ce{H2CO} $J$=3$_{0,3}$--2$_{0,2}$ gas mass ($M_{\rm \ce{H2CO}}$, \textit{center column}), and \ce{c-C3H2} $J$=3$_{3,0}$--2$_{2,1}$ gas mass ($M_{\rm \ce{c-C3H2}}$, \textit{right column}) versus envelope dust mass ($M_{\rm env}$) for the sample. In all panels the red dash-dotted line shows the \ce{N2H+} gas mass versus envelope dust mass from Paper I as reference. The full sample with no differentiation for multiplicity is shown in the top row. In the other three rows, orange stars indicate multiple protostellar systems, cyan diamonds show binary systems, and gray circles indicate single protostellar systems. Each row represents one of four ways of grouping the sample and their corresponding correlations (see Sect.~\ref{subsec:stats}). Lines and shaded areas show the linear regression for the data with the corresponding color. Solid lines indicate statistically significant correlations in the subsamples (Censored Kendall rank correlation p-value $<$ 0.05), while the dashed line shows subsamples with p-values $>$ 0.05.}
		\label{fig:Mass_gas_dust_schemes}
	\end{figure*}
	
	\section{Results}
	\label{sec:results}
	
	In this work, we focus on three transitions of \ce{DCO+} and \ce{c-C3H2}, and four transitions of \ce{H2CO}.
	Transitions, frequencies and beam sizes for each molecular species are listed in Table~\ref{tab:lines} along with the range of line peaks, RMS noise levels and line widths for each species and transition.
	All the measured and derived values from the data used in this work will be made available online at CDS.
	
	To assure that the results presented in this work are not biased due to sample selection, the rate of detections per molecular transition was counted.
	Detection fractions were examined relative to the full sample, system multiplicity, region clustering, and system parameters (bolometric luminosity $L_{\rm bol}$ and envelope dust mass $M_{\rm env}$).
	
	In general, detection rates are inversely proportional to frequency regardless of the sample binning used. 
	For the sample in our work, \ce{p-H2CO} $J$=3$_{0,3}$--2$_{0,2}$ is detected in 92\% of systems, while $J$=3$_{2,2}$--2$_{2,1}$ and $J$=3$_{2,1}$--2$_{2,0}$ are only detected in less than 40\% of systems. 
	Two transitions of \ce{c-C3H2}, $J$=3$_{3,0}$--2$_{2,1}$ and $J$=6--5, are detected with a rate above 70\%, while the $J$=5$_{1,4}$--4$_{2,3}$ transition is detected in 59\% of systems.
	Similarly, \ce{DCO+} $J$=3--2 is detected in 92\% of the full sample, while the $J$=5--4 transition is detected in 60\% of systems.
	The \ce{DCO+} $J$=1--0 transition is detected in all sources located within the OTF maps.
	
	When considering detection rates versus multiplicity, sources show similar detection rates for the low transitions, while close binaries and single systems show similar detection frequencies for the higher transitions, which are typically about 20\% lower than the multiple systems.
	Region clustering and $L_{\rm bol}$ does not affect detection frequencies.
	On the other hand, detection frequencies decrease proportionately with $M_{\rm env}$.
	This is expected as lower mass leads to lower fluxes.
	Sources with $M_{\rm env}$ $\geq$ 1 M$_{\odot}$ have 100\% detection frequencies in most transitions (except \ce{H2CO} $J$=3$_{2,2}$--2$_{2,1}$ and $J$=3$_{2,1}$--2$_{2,0}$, with $\sim$60 and $\sim$80 \% detection frequencies, respectively). 
	For sources with $M_{\rm env}$ $<$ 0.1 M$_{\odot}$, the detection frequency drops to 33\% in the lowest transition of APEX observations, and no detection for higher transitions.

	\section{Analysis} 
	\label{sec:analysis}
	
	\subsection{Gas kinetic temperature}
	\label{subsec:Tgas}
		
	The availability of two or more transitions per molecular species in the APEX data enables estimates of gas kinetic temperature through the use of line peak ratios.
	The 218 and 360 GHz data have spatial resolutions of 28.7$\arcsec$ and 18$\arcsec$, respectively (see Sect.~\ref{subsec:APEXobs}). To calculate line ratios within a common area of 28.7$\arcsec$, a beam dilution factor of 0.39 was applied to the \ce{DCO+} $J$=5--4 and \ce{H2CO} $J$=5$_{0,5}$--4$_{0,4}$ transitions.
	Applying the above beam dilution factor assumes that each line emission arises from a region that has the size of the corresponding beam, which might not be true for some sources or transitions. 
	However, to determine the relative extent of each species and transition, higher spatial resolution data is needed.
	We then calculate line peak ratios using the peak brightness temperature from the APEX observations for the following transition pairs: \ce{DCO+} $J$=5--4 / $J$=3--2; \ce{H2CO} $J$=3$_{2,2}$--2$_{2,1}$ / $J$=3$_{0,3}$--2$_{0,2}$, $J$=3$_{2,1}$--2$_{2,0}$ / $J$=3$_{0,3}$--2$_{0,2}$, and $J$=5$_{0,5}$--4$_{0,4}$ / $J$=3$_{0,3}$--2$_{0,2}$; and \ce{c-C3H2} $J$=6--5 / $J$=3$_{3,0}$--2$_{2,1}$ and $J$=5$_{1,4}$--4$_{2,3}$ / $J$=3$_{3,0}$--2$_{2,1}$.
	Only ratios where the lower transition is detected with $\geq$3$\sigma$ are considered. If the higher transition is not detected, the rms noise level is adopted for that transition and the ratio is considered an upper limit.
	Propagation of uncertainty analysis was used to calculate the line peak ratio error, using the noise level of the respective spectra as the standard deviation and assuming a covariance of zero (uncorrelated variables).
	
	Non-local thermal equilibrium (non-LTE) excitation and radiative transfer calculations using the off-line version of the program RADEX \citep{vandertak2007} with uniform sphere geometry together with molecular collisional data from the Leiden Atomic and Molecular Database (LAMDA, \citealt{schoier2005,vandertak2020}) were used to model the line peak ratios of \ce{DCO+}, \ce{H2CO} and \ce{c-C3H2}.
	We assume that the emission from the different transitions of the same molecular species arise from the same region, hence assuming that the material traced by the molecular species has a single density and temperature.
	Given that the Perseus molecular cloud dominates the emission along the line of sight (e.g., \citealt{doi2021}), we assume that there is no significant temperature gradient along the line of sight.
	The collisional rate coefficients for \ce{DCO+} are based on \citet{botschwina1993} and \citet{flower1999}, while for \ce{H2CO} and \ce{c-C3H2} the collisional rate coefficients are based on \citet{wiesenfeld2013} and \citet{chandra2000}, respectively.
	The line peak ratio models consider a temperature range of 10 to 200 K, and a \ce{H2} volume density range of 10$^{3}$ to 10$^{10}$.
	Column densities of 10$^{12}$ cm$^{-2}$ and line widths of 1 km~s$^{-1}$ were adopted in the models. The chosen line width is based on the average line width of all detected species, with broader line widths likely to have contributions from outflow emission (Fig.~\ref{fig:APEX_spec}).
	
	Gas kinetic temperatures $T_{kin}$ are inferred by comparing the observed line peak ratios and known \ce{H2} volume densities with the RADEX models.
	The \ce{H2} volumetric densities from Paper I were derived from the \ce{HNC} $J$=4--3 / $J$=1--0 ratio and are mean values within $\sim$5000 AU diameter regions, ranging from 10$^5$ to 10$^6$ cm$^{-3}$.
	Increasing the adopted \ce{H2} volumetric density by an order of magnitude decreases the estimated $T_{kin}$ by 5 K, at most, for all three molecular species used in this work.
	On the other hand, increasing or decreasing by an order of magnitude the adopted column density of a species in the RADEX models produces an average change of 2 K at most.
	The $T_{kin}$ adopted in this work were estimated from the APEX line peak ratios and the Paper I \ce{H2} volume density values, and their respective uncertainties, without corrections for the 28$\arcsec$ beamsize ($\sim$8000 AU) of the APEX observations.
	This comparison resulted in a list of values for which the average and standard deviation were calculated in order to adopt a single $T_{kin}$ value and uncertainty for each line peak ratio.
	
	The $T_{kin}$ values obtained from the three line peak ratios of \ce{H2CO} vary by about 10 to 20 K. This variation is likely caused by the variety of structures contributing to the emission within the APEX beam. However, analyzing the individual $T_{kin}$ values separately without spatially resolved maps of the emission would not provide deeper insight into the physical condition of each core.
	Hence, to avoid analyzing three separate $T_{kin}$ values obtained from the three \ce{H2CO} line peak ratios, the values were averaged together and likely contributes to the spread in the \ce{H2CO} $T_{kin}$ values reported in this work.
	In the case of \ce{c-C3H2}, the resulting $T_{kin}$ values from the two line peak ratios differ by 5 K or less, suggesting a common emission source for the three transitions. 
	Hence, the two separate $T_{kin}$ values were averaged together.
	
	The inferred $T_{kin}$ are plotted against $L_{\rm bol}$ and $M_{\rm env}$ in Fig.~\ref{fig:TN_LM}.
	Gas kinetic temperatures derived from \ce{DCO+} and \ce{c-C3H2} show similar values between 10 and 26 K. 
	These values are consistent with $T_{kin, \ce{NH3}}$ estimates toward Perseus from \citet{rosolowsky2008}, who found an average temperature of 11 K with a spread of 9 to 26 K from \ce{NH3} single pointing observations with the Green Bank Telescope (beam 31$\arcsec$), and $T_{kin}$ from $I$(\ce{HCN})/$I$(\ce{HNC}) $J$=1--0 maps (Paper I), which has an average value of 15 K.
	By-eye coordinate matching was used to find a common sample between the \citet{rosolowsky2008} catalog and our sample. The $T_{kin, \ce{NH3}}$ values for the common sample are plotted in Fig.~\ref{fig:TN_LM} for comparison with the $T_{kin}$ values derived in this work.
	On the other hand, \ce{H2CO} ranges between 10 and 40 K, with one source showing a gas kinetic temperature of 80 K (NGC1333 IRAS5 Per63, outside the plotted range of Fig.~\ref{fig:TN_LM}) and another with an upper limit of 54 K (L1455 Per25).

	\subsection{Column density}
	\label{subsec:colden}
	
	Using the observed peak brightness temperature and line width together with the \ce{H2} volumetric density (n(\ce{H2})) from Paper I, and the gas kinetic temperature ($T_{\rm kin}$) derived in this work (Sect.~\ref{subsec:Tgas}), we derive the column density of each molecular species and transition from non-LTE calculations.
	The column density uncertainties were calculated by using the lower and upper limits of n(\ce{H2}) and $T_{\rm kin}$ in the non-LTE RADEX calculations, and then taking the average and standard deviation of all three values.
	Column densities for the different transitions of the same molecular species are found to be similar, varying by up to 50\%, with only a few cases varying by a factor of two.
	The column density for each transition was calculated in several ways: using two values of $T_{\rm kin}$ (derived from $I$(\ce{HCN})/$I$(\ce{HNC}) $J$=1--0 or corresponding line ratio) and varying n(\ce{H2}) by an order of magnitude. 
	The objective being to determine how much does the choice of $T_{\rm kin}$ and n(\ce{H2}) affect the calculated column density.
	For all molecular species, upper limits lead to severely under- or overestimated column densities.
	
	The column density of \ce{DCO+} is not sensitive to the adopted gas kinetic temperature.
	Increasing n(\ce{H2}) by an order of magnitude causes N(\ce{DCO+}) to increase by a factor of 2 on average for $J$=3--2 and a factor of 4 on average for $J$=5--4.
	In the case of \ce{H2CO}, increasing $T_{\rm kin}$ and n(\ce{H2}) typically generates lower N(\ce{H2CO}) column densities by a few percent.
	For \ce{c-C3H2}, changing $T_{\rm kin}$ does not cause significant changes in N(\ce{c-C3H2}).
	Overall, the calculated column densities do not vary by one or more orders of magnitude unless one of the parameters is an upper limit.
	
	The calculated column densities used in this work adopt the n(\ce{H2}) from Paper I, and the $T_{\rm kin}$ from the corresponding molecular species (see Sect.~\ref{subsec:Tgas}).
	For \ce{DCO+} $J$=1--0, the $T_{\rm kin}$ from Paper I was adopted instead of the gas kinetic temperature derived from the $J$=5--4 / $J$=3--2 ratio.
	All derived column densities are available in the CDS published data.
	The column densities for the lowest transition of \ce{DCO+}, \ce{H2CO} and \ce{c-C3H2} are plotted versus $L_{\rm bol}$ and $M_{\rm env}$ in Fig.~\ref{fig:TN_LM}.

	\subsection{Gas mass}
	\label{subsec:Mgas}
	
	Different molecular species trace distinct temperature regimes and various physical structures.
	\ce{DCO+} traces the cold inner envelope where gaseous \ce{CO} starts to freeze out. \ce{H2CO} traces the inner warm envelope and outflow cavity. \ce{c-C3H2} traces the irradiated outflow cavity. In addition, not all sources present emission in all three species and their respective transitions (see Sect.~\ref{sec:results}). By calculating the gas mass of the three species, we can probe the amount and distribution of material available for the forming protostar and compare this distribution with multiplicity. 
	
	The gas mass for a molecular species X, $M_{\rm gas, X}$, was derived in the same way as in Paper I, using the expression
	\begin{equation} 
		\label{eg:mass}
		M_{\rm gas, X} = \frac{M_{r}~(\rm g~mol^{-1}) N(X)~(\rm cm^{-2})~A~(\rm cm^2)}{N_{A}~(\rm mol^{-1})~M_{sun,g} ~(\rm g)} ~~M_{\odot}
	\end{equation}
	where $M_{r}$ is the mean molecular weight (30.0237, 30.031, 38.01565 g~mol$^{-1}$ for \ce{DCO+}, \ce{H2CO}, \ce{c-C3H2}, respectively), $N_{A}$ is Avogadro's constant, $N(X)$ is the derived gas column density for $X$ molecule, $A$ is the beam area over which the mass is calculated, and $M_{sun, g}$ is the solar mass in grams.
	Adopted distances per subregion are 288 pc for L1448, 299 pc for NGC1333 and L1455, 301 pc for B1, and 295 pc for IC348 \citep{zucker2018}.
	The beam diameters are listed in Table~\ref{tab:lines}.
	The uncertainty is obtained with the same equation, but with $N(X)$ being the uncertainty of the derived column density and all other parameters assumed to be constant.
	The derived $M_{\rm gas, X}$ are plotted versus $M_{\rm env}$ in Fig.~\ref{fig:Mass_fullSample}.
	All derived gas masses are available in the CDS published data. 
	Derived gas masses are consistent between transitions, showing similar values and trends, in particular the trend of gas mass increasing proportionately to $M_{\rm env}$. 
	The gas mass obtained from \ce{H2CO} shows a larger spread than the gas masses from \ce{DCO+} and \ce{c-C3H2}.

	\subsection{Statistical analysis}
	\label{subsec:stats}
	
	To examine the relations between the physical parameters derived in this work and the multiplicity of the sample, a statistical correlation test is performed.
	The Generalized Kendall's rank correlation \citep{isobe1986} is applied to our data using standard python libraries and following the method used in the Pilot Paper.
	This method enables the degree of association between two quantities with censored (upper limit) data to be measured.
	For this work, $L_{\rm bol}$ and $M_{\rm env}$ were tested against gas kinetic temperature and gas mass.
	The null hypothesis of the Generalized Kendall's rank correlation is that the values are uncorrelated.
	Hence, p-values $<$0.05 indicate a correlation at better than 3$\sigma$ significance.
	
	Four classification schemes, listed in Table~\ref{tab:schemes}, are used for statistical analysis.
	These classification schemes follow those used in Paper I, except for multiplcity within beam (Scheme 2 in Paper I) which is not followed due to the different angular resolutions of the APEX and NRO data (Table~\ref{tab:lines}).
	Bin sizes in each classification scheme vary by molecular species used to derive the physical parameter being tested. This occurs because not all sources have detections in all transitions of each molecular species.
	A system in this work is considered to be all the protostars within a common core, assuming gravitational boundness for simplicity.
	This definition leads to gas and dust mass of a system being equal to the sum of masses of each of its components, while gas kinetic temperature and column densities are averaged over all components.
	
	Gas kinetic temperatures do not show a correlation to $L_{\rm bol}$ regardless of the classification scheme. 
	The Generalized Kendall's rank correlation only finds p-values $<$ 0.05 between $M_{\rm env}$ and $T_{\rm kin}$ from \ce{DCO+} $J$=5--4 / $J$=3--2 for the full sample. In the other three classification schemes, the correlation only appears in the single protostar bin.
	The reason for this result is unclear, and may require data that spatially resolves the molecular line emission to explain.
	As expected from Fig.~\ref{fig:TN_LM}, the slope of the $T_{\rm kin}$ versus $L_{\rm bol}$ or $M_{\rm env}$ is flat.

	Gas mass $M_{gas, X}$ from the three molecular species is uncorrelated to $L_{\rm bol}$ in any of the classification schemes.
	In contrast, $M_{gas, X}$ from the three molecular species studied shows correlation to $M_{\rm env}$ in the four classification schemes.

	In Fig.~\ref{fig:Mass_gas_dust_schemes} the general trend from $M_{\rm gas, \ce{N2H+}}$\footnote{In the CDS published data, $N$(\ce{N2H+}), $N$(\ce{HCO+}), $M_{\rm gas, \ce{N2H+}}$ and $M_{\rm gas, \ce{HCO+}}$ from Paper I are also included with their corresponding uncertainties. In Paper I, $N$(\ce{HCO+}) was erroneously reported at a factor of 16 lower than $N$(\ce{N2H+}). Instead, $N$(\ce{HCO+}) is a factor of 50 lower than $N$(\ce{N2H+}).} reported in Paper I is shown for reference and comparison.
	All gas masses derived in this work are about two orders of magnitude lower than the $M_{\rm gas, \ce{N2H+}}$ but with a similar slope. 
	$M_{\rm gas, \ce{DCO+}}$, $M_{\rm gas, \ce{H2CO}}$ and $M_{\rm gas, \ce{c-C3H2}}$ are similar to $M_{\rm gas, \ce{HCO+}}$.
	This is likely produced by the molecular species and transitions used in this work tracing the inner regions of the protostellar cores, which naturally have less material within the beam than \ce{N2H+}.
	It is unlikely to stem from chemical processes (e.g., fractionation) due to the lack of correlation between $M_{gas, X}$ and $T_{\rm kin}$ or $L_{\rm bol}$. 
	
	In classification schemes 2 and 4, $M_{\rm gas, \ce{DCO+}}$ and $M_{\rm env}$ exhibits a correlation for all multiplicity bins, with multiple systems showing slightly higher masses than singles and binaries (Fig.~\ref{fig:Mass_gas_dust_schemes}, left column).
	In classification scheme 3, the multiple systems do not show a significant correlation between $M_{\rm env}$ and $M_{\rm gas, \ce{DCO+}}$ or $M_{\rm gas, \ce{c-C3H2}}$ which may be due to the bin sample size of five and four systems.
	For \ce{H2CO} (Fig.~\ref{fig:Mass_gas_dust_schemes}, center column), classification schemes 2 and 3 do not show significant correlations between multiple systems and mass (bin sample sizes of 10 and 5, respectively). 
	Binaries in classification scheme 2 (bin sample size of 7) also do not show a significant correlation between $M_{\rm gas, \ce{H2CO}}$ and $M_{\rm env}$.
	The spread in values derived from \ce{H2CO} may also contribute to weaken any correlation if it exists.
	Finally, $M_{\rm gas, \ce{c-C3H2}}$ (Fig.~\ref{fig:Mass_gas_dust_schemes}, right column) presents less spread than $M_{\rm gas, \ce{H2CO}}$.
	Single systems in classification schemes 2 to 4 do not show correlations between $M_{\rm gas, \ce{c-C3H2}}$ and $M_{\rm env}$, while binary systems do not show $M_{\rm gas, \ce{c-C3H2}}$ and $M_{\rm env}$ correlation in classification scheme 2.
	Multiple systems show correlations in classification schemes 2 and 4 (bin sample sizes of 9 and 14, respectively) and generally higher $M_{\rm gas, \ce{c-C3H2}}$ and $M_{\rm env}$ than singles and binaries.

	\section{Discussion}
	\label{sec:discussion}
	
	\subsection{Physical parameters}
	
	Gas kinetic temperature obtained from all three molecular species presented in this work is uncorrelated to bolometric luminosity $L_{\rm bol}$ or envelope dust mass $M_{\rm env}$ (Fig.~\ref{fig:TN_LM}), consistent with the results from Paper I.
	For the full sample, $T_{\rm kin, \ce{DCO+}}$ and $T_{\rm kin, \ce{c-C3H2}}$ have average values of 14 and 16 K, respectively, spreading between 10 to 26 K.
	This is similar to the results from $I$(\ce{HCN})/$I$(\ce{HNC}) $J$=1--0, an average of 15 K and spread of 15 -- 20 K (Paper I).
	$T_{\rm kin, \ce{H2CO}}$ has a higher average value of 26 K, with a much larger spread in gas kinetic temperatures (13 -- 82 K).
	
	While \ce{c-C3H2} is typically associated with outflow cavities (e.g., \citealt{tychoniec2021}), the average $T_{\rm kin, \ce{c-C3H2}}$ in this work does not change if only the smaller ($\sim$8300 AU at $d$=288 pc) or larger ($\sim$8600 AU at $d$ = 301 pc) areas are considered. 
	For $T_{\rm kin, \ce{DCO+}}$ and $T_{\rm kin, \ce{H2CO}}$, the average will change by 3 K and 1 K, respectively, between smaller and larger distances.
	These results underline the fact that protostellar heating above 30 K is concentrated in small regions ($<<$1000 AU) around the protostar.
	This has been demonstrated in previous observational and modeling work \citep{harsono2011,krumholz2014,murillo2016,friesen2017,offner2022ppvii,mignon-risse2021}.
	Furthermore, most of the heating from protostars has been shown to escape mainly through the outflow cavity \citep{yildiz2015,guszejnov2021,mathew2021} and is diluted in the beam of the APEX observations.

	Gas masses $M_{\rm gas, X}$ show statistically significant correlations with $M_{\rm env}$ for all three molecular species studied in this work (Sect.~\ref{subsec:stats}).
	$M_{\rm gas, X}$ calculated for each transition of the same molecular species are quite consistent (Fig.~\ref{fig:Mass_fullSample}, Sect.~\ref{subsec:Mgas}).
	All three molecular species show $M_{\rm gas, X}$ lower by about two orders of magnitude than $M_{\rm gas, \ce{N2H+}}$ (Paper I), which is consistent with the chemistry of \ce{DCO+}, \ce{c-C3H2} and \ce{H2CO}.
	Three trends are found for all classification schemes, which are seen more clearly in \ce{DCO+} and \ce{c-C3H2} than in \ce{H2CO}.
	Multiple systems tend to have slightly higher $M_{\rm gas, X}$ and $M_{\rm env}$ than single and binary systems. 
	In contrast, binaries and single protostellar systems show similar $M_{\rm gas, X}$ and $M_{\rm env}$. 
	
	A continuous $M_{\rm gas, X}$ and $M_{\rm env}$ distribution is observed for binaries and higher-order multiple systems, with no differentiation for small ($\leq$7$\arcsec$) and large ($>$7$\arcsec$) separations (see Fig.~\ref{fig:Mass_gas_dust_schemes}, second and fourth rows). 
	This lack of differentiation between small and large separations is worth considering in the context of binary and higher-order multiple protostellar system formation pathways. Traditionally, theory has supported disk fragmentation as the mechanism that produces small separations ($<$100 AU), while core fragmentation produces larger separations (see \citealt{offner2022ppvii} and references therein). \citet{tobin2016} reported a bimodal distribution in separations (peaks at 75 and 3000 AU) of protostellar components from the VANDAM survey toward Perseus. The proposed explanation for the bimodal distribution is the difference in formation mechanisms, as well as the possibility of inspiral motion of components in the protostellar system.
	Instead, \citet{kuruwita2023} find that separations between 20 to 100 AU can be reproduced through core fragmentation or dynamical capture with efficient inspiraling of protostellar components. 
	The continuous gas and dust mass distributions with no differentiation for separation would further support the results from \citet{kuruwita2023} which indicate that separate formation pathways are not necessary to reproduce the different separations observed in dust continuum surveys.
	
	The spread in mass and temperature derived from \ce{H2CO} using non-LTE assumptions (Fig.~\ref{fig:TN_LM} and \ref{fig:Mass_fullSample}) is likely due to the broad range of structures this molecule can trace, from inner envelope to outflow cavities (e.g., \citealt{tychoniec2021}).
	Hence, the \ce{H2CO} derived parameters need to be considered with caution.
	Observations which spatially resolve the \ce{H2CO} emission are needed to fully understand the derived parameters.
	In the same vein, it would be interesting to explore how the spatial distribution of mass within cloud cores affects the hierarchy of higher-order multiple protostellar systems.
	
	\subsection{Filament hubs and magnetic fields}
	
	The results of the current work and Paper I indicate that large reservoirs of gas and dust mass are key for higher-order multiple system formation.
	On the other hand, binaries (regardless of separation) and single sources can form from protostellar cloud cores of a few Jeans masses, as has been pointed out in previous work (e.g., \citealt{pon2011,mairs2016,johnstone2017}).
	There must then be an additional mechanism that enables a cloud core to have access to large mass reservoirs.
	Studies on gas kinematics and structure in the Perseus molecular cloud could provide insight into this question.
	Observations of B1, both the main core (where B1-a, B1-b, B1-c, B1 Per6+Per10 are located) and the ridge (where IRAS 03292+3039 and IRAS 03282+3035 are located), suggest that the main core is more kinematically complex than the ridge \citep{storm2014}.
	Protostellar cores show complex kinematics that cannot be explained solely with infall or rotation \citep{chen2019}.
	Some studies suggest that the filamentary structure of molecular clouds could arise before protostars form \cite{storm2016} or be product of turbulent cells that collide with the cloud and compress the gas \citep{dhabal2019}.
	Recent observations with IRAM 30m and NOEMA, tracing scales down to $>$1400 AU and covering the region between NGC1333 IRAS4 and SVS13, suggest that streamers could deliver gas from the molecular cloud to the inner region of protostellar cores \citep{valdivia2024}.
	Using data from the Green Bank Ammonia Survey \citep{friesen2017} toward several Perseus molecular cloud regions, \citet{chen2024} find that the velocity dispersion and filament mass per unit length correlate with number of cores and embedded protostars. At the same time, \citet{chen2024} highlight the complex kinematic structures in their data and at smaller scales ($<$0.02 pc).
	This is consistent with previous studies, both toward Perseus, as mentioned above, and regions like Serpens where a combination of gravitational infall and collapse within filaments was observed \citep{fernandez2014}.
	The analytical model of \citet{hacar2024} suggests that the efficient merging of filaments can lead to cores forming at the junctions of filaments with high accretion rates.
	These cores collapse and gain mass from the filament hub while continuing to fragment.
	\citet{hacar2024} note that the fractal nature of the ISM implies that this mechanism is scale free.
	If higher-order multiple protostellar systems form at the junction of filament hubs, this would provide an increased mass reservoir (and angular momentum, e.g., \citealt{boss1995}) available for continued fragmentation and mass accretion.
	However, given the complexity of the kinematics in the filaments \citep{fernandez2014,storm2014,chen2024} and at core scales \citep{chen2019,valdivia2024}, not all protostellar cores would receive the same delivery and distribution of material.
	Indeed, multiple protostellar systems have been observationally shown to have non-coeval components (protostars at different evolutionary stages within a common core, e.g., \citealt{murillo2016,luo2022}). This phenomena could be product of continued accretion of material from larger scales, leading to further fragmentation, or caused by dynamical evolution (e.g., \citealt{kuruwita2023}).

	Magnetic fields could also be a factor in multiple star formation, as has been explored in previous work \citep{padoan2014,mignon-risse2021}. 
	These studies found that magnetic pressure dominates thermal pressure at the core scale, which is supported by the results of our work. 
	Studies of magnetic field morphology and strength toward sources in Orion indicate that standard hourglass morphology is related to larger field strengths, while other morphologies (e.g., spiral, rotated hourglass, complex) are related to weaker field strengths (\citealt{huang2024a,huang2025}). 
	Considering the field morphology of the multiple protostellar systems included in the sample of \citet{huang2024a}, a tendency for weaker magnetic fields strengths toward multiple systems may be a possible scenario.
	However, there is a considerable uncertainty in deriving the true magnetic field. 
	It is also unclear how the magnetic field strength would contribute to the mass reservoir for a protostellar cloud core.
	There could be a possible scenario where both parameters (mass and magnetization) would, independently, matter.
	
	\section{Conclusions}
	\label{sec:conclusions}
	
	In this paper, we present Atacama Pathfinder EXperiment (APEX) and Nobeyama 45m Radio Observatory (NRO) observations covering five subregions in the Perseus molecular cloud. The observations targeted a sample of 31 protostellar systems.
	Several transitions of three molecular species, \ce{DCO+}, \ce{H2CO} and \ce{c-C3H2}, were targeted and used to calculate gas kinetic temperature $T_{\rm kin, X}$, column density $N(X)$, and gas mass $M_{\rm gas, X}$.
	These parameters have been proposed from theory, models or previous observational studies (e.g., Pilot paper and Paper I, and references therein) to be key in determining the multiplicity of low-mass protostellar systems.
	
	The results reported in this work confirm the findings of the Pilot Paper and Paper I, namely:
	\begin{enumerate}
		\item Mass is a key factor in determining protostellar multiplicity, especially for higher-order multiple protostellar systems.
		\item A continuum in the formation mechanisms, rather than distinct formation mechanisms for close and wide separations, is suggested based on the continuous gas and dust mass relations found for all molecular species.
		\item Gas kinetic temperature $T_{\rm kin}$ is not related to protostellar multiplicity.
	\end{enumerate}
	How the gas mass is spatially distributed within the protostellar cloud core at scales below 5000 AU must be explored in future work. 
	The factors that enable or hinder a cloud core to accumulate mass to eventually form higher-order multiple protostellar systems will also need to be further analyzed.
	The most intriguing question then becomes how does the mass accumulation of cloud cores affect the gravitational boundness of higher-order multiple systems, their eventual evolution, and survivability.
	
	\textit{Data availability:} The source information and corresponding physical parameters derived in this work, listed in Table~\ref{tab:source} and plotted in the figures, is available at the CDS via anonymous ftp to \url{cdsarc.u-strasbg.fr} (130.79.128.5) or via \url{http://cdsweb.u-strasbg.fr/cgi-bin/qcat?J/A+A/}.

	\begin{acknowledgements}
		This paper made use of Nobeyama data, and APEX data.
		The Nobeyama 45-m radio telescope is operated by Nobeyama Radio Observatory, a branch of National Astronomical Observatory of Japan. 
		We are grateful to the APEX staff for support with these observations. Observing time for the APEX data was obtained via Max Planck Institute for Radio Astronomy, Onsala Space Observatory and European Southern Observatory.
		The authors acknowledge the use of the Canadian Advanced Network for Astronomy Research (CANFAR) Science Platform operated by the Canadian Astronomy Data Centre (CADC) and the Digital Research Alliance of Canada (DRAC), with support from the National Research Council of Canada (NRC), the Canadian Space Agency (CSA), CANARIE, and the Canada Foundation for Innovation (CFI).
		This study is supported by a grant-in-aid from the Ministry of Education, Culture, Sports, Science, and Technology of Japan (20H05645, 20H05845 and 20H05844) and by a pioneering project in RIKEN Evolution of Matter in the Universe (r-EMU).
		D.H. is supported by a Center for Informatics and Computation in
		Astronomy (CICA) grant and grant number 110J0353I9 from the Ministry
		of Education of Taiwan. D.H. also acknowledges support from the
		National Science and Technology Council, Taiwan (Grant
		NSTC111-2112-M-007-014-MY3, NSTC113-2639-M-A49-002-ASP, and
		NSTC113-2112-M-007-027).
		D.J. is supported by NRC Canada and by an NSERC Discovery Grant.
		R.M.R. has received funding from the European Research Concil (ERC) under the European Union Horizon 2020 research and innovation programme (grant agreement number No. 101002352, PI: M. Linares)
	\end{acknowledgements}

	\bibliographystyle{aa}
	\bibliography{aa54308-25.bib}

	\begin{appendix}
	\section{Sample}
	Table~\ref{tab:source} lists the source sample discussed in this work and described in Sect.~\ref{subsec:sample}. Table~\ref{tab:source} lists the system names, right ascension (R.A.) and declination (Dec.) coordinates, projected separation from a reference component, whether they are located in a clustered environment or not, and the bolometric luminosity and envelope dust mass adopted in the current work. The latter two parameters were calculated according to the method described in Sect.~\ref{subsec:dustcontparam}.
	
	    \begin{table*} 
		\centering
		\caption{Sample of protostellar systems}
		\begin{tabular}{c c c c c c c c}
			\hline \hline
			System & Sources & R.A. & Dec. & Projected  & Clustered?\tablefootmark{a} & $L_{\rm bol}$ (L$_{\odot}$) & $M_{\rm env}$ (M$_{\odot}$) \\
			&  &  &  & Separation ($\arcsec$) &  & & \\
			\hline
			\multicolumn{8}{c}{Wide multiples}\\
			\hline
			L1448 N\tablefootmark{b}  &  A  &  03:25:36.53  &  30:45:21.35  &  …  &  N  & 12.38 & 2.73 \\
			&  B  &  03:25:36.34  &  30:45:14.94  &  7.3  &    & 3.23 & 4.43 \\
			&  C  &  03:25:35.53  &  30:45:34.20  &  16.3  &    & 4.43 & 1.35 \\
			L1448 C  &  N  &  03:25:38.87  &  30:44:05.40  &  …  &  N  & 1.44 & 2.04 \\
			&  S  &  03:25:39.14  &  30:43:58.30  &  8.1  &    & 2.54 & 1.67 \\
			NGC1333 IRAS7\tablefootmark{b}  &    PER18  &  03:29:11.26  &  31:18:31.08  &  …  &  Y  & 7.72 & 0.71 \\
			&    PER21  &  03:29:10.67  &  31:18:20.18  &  13.3  &    & 5.67 & 0.73 \\
			&    PER49  &  03:29:12.96  &  31:18:14.31  &  27.5  &    & 1.05 & 0.40 \\
			NGC1333 IRAS4B\tablefootmark{c} &  B  &  03:29:12.01  &  31:13:08.10  &  …  &  Y  & 14.15 & 2.53 \\
			&  B’  &  03:29:12.83  &  31:13:06.90  &  10.6  &    & 0.03 & 21.34 \\
			NGC1333 SVS13\tablefootmark{b}  &  A  &  03:29:03.75  &  31:16:03.76  &  …  &  Y  & 190.17 & 0.67 \\
			&  B  &  03:29:03.07  &  31:15:52.02  &  14.9  &    & 16.61 & 1.56 \\
			&  C  &  03:29:01.96  &  31:15:38.26  &  34.7  &    & 3.66 & 1.15 \\
			NGC1333 IRAS2  &  A  &  03:28:55.57  &  31:14:37.22  &  …  &  Y  & 76.26 & 0.86 \\
			&  B  &  03:28:57.35  &  31:14:15.93  &  31.4  &    & 9.54 & 0.30 \\
			B1-b\tablefootmark{b}  &  S  &  03:33:21.30  &  31:07:27.40  &  …  &  N  & 0.57 & 2.98 \\
			&  N  &  03:33:21.20  &  31:07:44.20  &  17.4  &    & 0.28 & 2.84 \\
			&  W  &  03:33:20.30  &  31:07:21.29  &  13.9  &    & 0.20 & 3.18 \\
			B1 P6+P10  &    Per6  &  03:33:14.40  &  31:07:10.88  &  …  &  N  & 0.34 & 0.72 \\
			&    Per10  &  03:33:16.45  &  31:06:52.49  &  31.9  &    & 0.77 & 0.76 \\
			IC348MMS  &    Per11  &  03:43:57.06  &  32:03:04.60  &  …  &  N  & 3.69 & 1.16 \\
			&  E  &  03:43:57.73  &  32:03:10.10  &  10.2  &    & 0.16 & 3.60 \\
			IC348SMM2  &  S  &  03:43:51.08  &  32:03:08.32  &  …  &  N  & 2.66 & 0.31 \\
			&  N  &  03:43:51.00  &  32:03:23.76  &  16.1  &    & 1.51 & 0.43 \\
			IC348 P32+ED366  &    Per32  &  03:44:02.40  &  32:02:04.89  &  …  &  N  & 0.27 & 0.47 \\
			&    ED366  &  03:43:59.44  &  32:01:53.99  &  36.6  &    & 2.05 & 0.09 \\
			\hline               
			\multicolumn{7}{c}{Close binaries}\\               
			\hline               
			Per17  &    &  03:27:39.09  &  30:13:03.00  &  0.273  &  N  & 13.74 & 0.23 \\
			L1455 FIR2  &    &  03:27:38.23  &  30:13:58.80  &  0.346  &  N  & 1.04 & 0.18 \\
			NGC1333 IRAS4A\tablefootmark{c} & A1 \ A2 &  03:29:10.51  &  31:13:31.01  &  1.8  &  Y  & 16.46 & 6.08 \\
			NGC1333 IRAS1\tablefootmark{b}  & N \ S &  03:28:37.00  &  31:13:27.00  &  1.9  &  Y  & 17.81 & 0.19 \\
			EDJ2009-156  &    &  03:28:51.11  &  31:18:15.41  &  3.192  &  Y  & 0.84 & 0.08 \\
			IRAS 03282+3035\tablefootmark{b,d}  &    &  03:31:21.00  &  30:45:30.00  &  0.098  &  N  & 2.44 & 0.84 \\
			IRAS 03292+3039  &    &  03:32:17.95  &  30:49:47.60  &  0.085  &  N  & 1.76 & 2.11 \\
			B1-a  &    &  03:33:16.66  &  31:07:55.20  &  0.391  &  N  & 2.30 & 0.22 \\
			\hline               
			\multicolumn{7}{c}{Single systems}\\               
			\hline               
			L1455 IRS4  &    &  03:27:43.23  &  30:12:28.80  &  …  &  N  & 2.83 & 0.28 \\
			L1455 Per25\tablefootmark{b}  &    &  03:26:37.46  &  30:15:28.01  &  …  &  N  & 1.76 & 0.27 \\
			L1448IRS2  &     &  03:25:22.40  &  30:45:12.00  &  …  &  N  & 6.50 & 0.78 \\
			L1448IRS2E   &      &  03:25:25.66  &  30:44:56.70  &  …  &  N  & 0.18 & 0.66 \\
			NGC1333 IRAS5\tablefootmark{b}  &    PER52  &  03:28:39.72  &  31:17:31.89  &  …  &  Y  & 0.23 & 0.44 \\
			NGC1333 IRAS5\tablefootmark{b}   &    PER63  &  03:28:43.28  &  31:17:32.90  &  …  &  Y  & 2.31 & 0.09 \\
			RNO 15 FIR  &    &  03:29:04.05  &  31:14:46.61  &  …  &  Y  & 0.47 & 0.97 \\
			NGC1333 SK1\tablefootmark{b}  &    &  03:29:00.52  &  31:12:00.68  &  …  &  Y  & 1.15 & 0.36 \\
			NGC1333 IRAS6  &    &  03:29:01.57  &  31:20:20.69  &  …  &  Y  & 19.49 & 0.44 \\
			NGC1333 IRAS4C\tablefootmark{c} &    &  03:29:13.52  &  31:13:58.01  &  …  &  Y  & 1.18 & 0.82 \\
			B1-c  &    &  03:33:17.85  &  31:09:32.00  &  …  &  N  & 8.50 & 1.41 \\
			HH211  &    &  03:43:56.80  &  32:00:50.21  &  …  &  N  & 2.54 & 1.43 \\
			
			\hline
		\end{tabular}
		\\
		\tablefoot{\tablefoottext{a}{N=No; Y=Yes. Clustered regions have 34 YSO~pc$^{-1}$, while non-clustered regions have 6 YSO~pc$^{-1}$ \citep{plunkett2013}.}
			\tablefoottext{b}{Systems from the sample in the Pilot Paper \citet{murillo2018PaperI}.}
			\tablefoottext{c}{Only included in the \ce{DCO+} $J$=1--0 Nobeyama data sample.}
			\tablefoottext{d}{\citet{reynolds2023} raise the possibility that IRAS 03282+3035 is not a close binary given recent observations, but cannot confirm. But given the results of the current work, there is no difference in classifying the system as a single or close binary in terms of physical parameters.}}
		\label{tab:source}
	\end{table*}	
		
	\section{Envelope Dust masses}
	\label{app:contdata}
	
	\begin{figure} 
		\centering
		\includegraphics[width=\columnwidth]{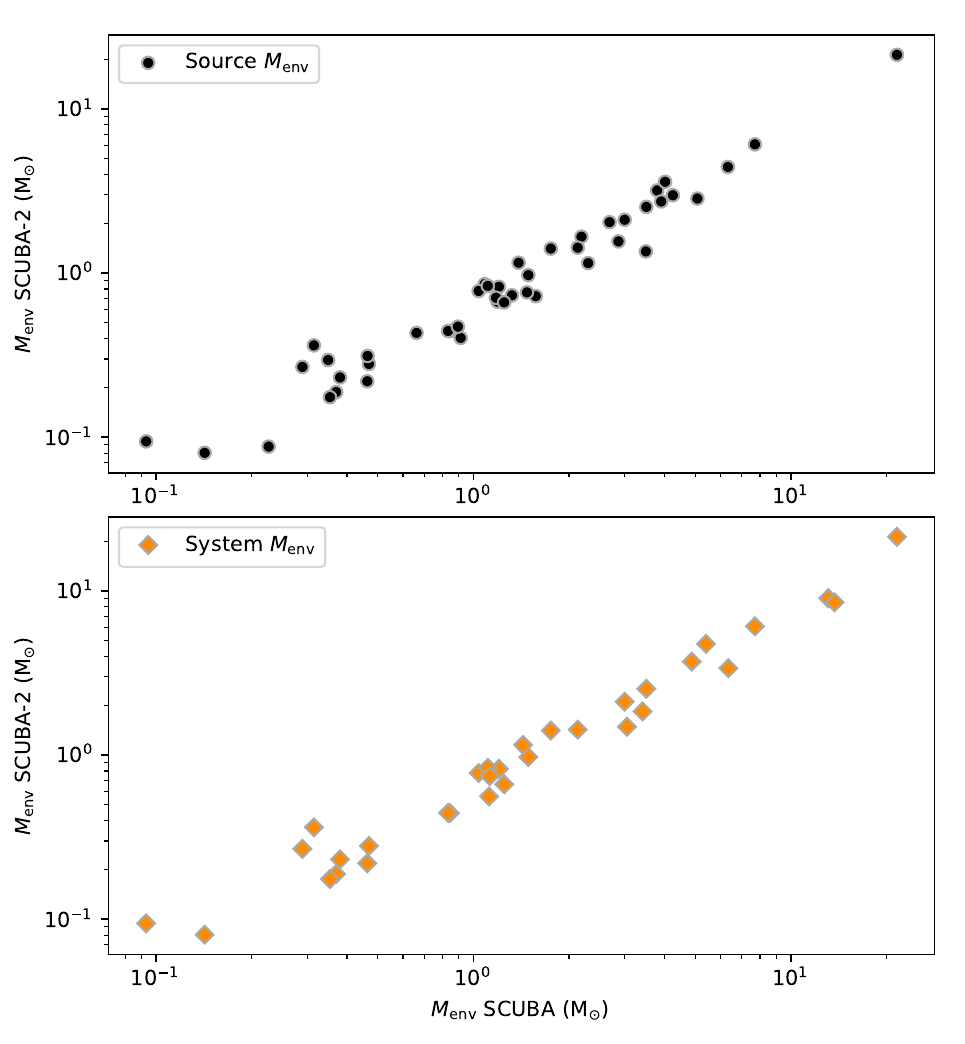}
		\caption{Comparison of envelope dust masses calculated with SCUBA (x-axis, COMPLETE survey, \citealt{ridge2006}) and SCUBA-2 (y-axis, JCMT GBS, \citealt{kirk2018}) data. The top panel shows the dust masses per source in our sample (i.e., APEX pointing), while the bottom panel shows the dust masses per system (i.e., summed masses of all components in multiple protostellar systems).}
		\label{fig:SCUBAmasses}
	\end{figure}
	
	This appendix shows the comparison of envelope dust masses derived from the SCUBA (COMPLETE survey, \citealt{ridge2006}) and SCUBA-2 (JCMT GBS\footnote{\url{https://doi.org/10.11570/18.0005}}, \citealt{kirk2018,chen2016}) 850$\mu$m maps. 
	The Pilot paper \citep{murillo2018PaperI} and Paper I \citep{paperII2024} used the SCUBA data to calculate envelope dust masses.
	For this work, we used the data release 3 (DR3) \ce{CO}-subtracted fits maps for Perseus West and IC348.
	The GBS DR3 maps use units of mJy arcsec$^{-2}$ and have a FWHM beam of 14.4$\arcsec$. The SCUBA maps use units of Jy~beam$^{-1}$ and have a beam of 15$\arcsec$. 
	In order to convert the SCUBA-2 peak fluxes from arcsec$^{2}$ to beam, we calculated the area of the beam as $A_{beam}$ = 2~$\pi$~$\sigma^2$, where $sigma$ = FWHM / 2 $\sqrt(2 ln(2))$ for a gaussian beam. For a FWHM of 14.4$\arcsec$ we used $A_{beam}$ $\eqsim$ 234.96 arcsec$^2$.
	Envelope dust masses were derived according to the method explained in Sect.~\ref{subsec:dustcontparam}.
	The calculation shows that the SCUBA-2 masses vary by factors of 0.9--2.6 when compared to the SCUBA masses (Fig.~\ref{fig:SCUBAmasses}). Despite this variation, the envelope dust mass trend does not change for the individual sources in our sample or for the masses of multiple protostellar systems.
	The calculated envelope dust masses from SCUBA and SCUBA-2 data are included in the CDS online data for the current work. The envelope dust masses listed in Table~\ref{tab:source} are those derived from SCUBA-2 peak fluxes.
	
	With the aim to further check that our envelope dust masses are reasonable, we compare our SCUBA calculated values with the dust masses from the core catalog of \citet{pezzuto2021}. 
	The core catalog was filtered to only consider cores classified as protostellar. By-eye coordinate matching was used to find a common sample of 35 cores for 31 systems in our sample. The match is not perfect since our work uses VANDAM survey positions \citep{tobin2016}, while \citet{pezzuto2021} used the \textit{getsources} algorithm on \textit{Herschel} PACS maps. This leads to different numbers of cores being associated to higher-order multiples in our sample (e.g., NGC1333 SVS13 A, B and C are associated to cores 314, 311 and 303 from \citet{pezzuto2021}; L1448 N A+B is associated to core 69, while L1448 N C is associated to core 68 from \citet{pezzuto2021}). 
	From the matched sample, we find that 21 out of 35 cores have masses differing by a factor of 3 or less. In addition, 24 out of 35 cores present higher envelope dust masses in our sample than in the \citet{pezzuto2021} core catalog.
	It should be noted that the core catalog employs spectral energy distribution (SED) fitting with a modified black body that is not suitable for protostellar sources \citep{pezzuto2021}, while we use a method dependent on $L_{bol}$, distance and peak flux at 850$\mu$m (see Sect.~\ref{subsec:dustcontparam} and \citealt{jorgensen2004}). The largest discrepancies come from the adopted dust temperature in the SED fitting method and the luminosity obtained from integrating the SED $L_{SED}$ \citep{pezzuto2021}.
	Core 311, corresponding to NGC1333 SVS13 B, does not have an SED fit in the core catalog. A dust temperature of 10.4 K (average of all reliable starless and prestellar SEDs) and $L_{SED}$ = 0.024 L$_{\odot}$ is adopted for Core 311. Whereas we adopt a dust temperature of 20.79 K \citep{zari2016} and $L_{bol}$ = 16.6 L$_{\odot}$ \citep{murillo2016} for SVS13 B. For this case, the dust mass from \citet{pezzuto2021} is 0.186 M$_{\odot}$, while our dust mass is 2.87 M$_{\odot}$, a factor of 15 difference and the largest discrepancy.	
		
	  \section{APEX spectra}
	  \label{app:spectra}
	  
\begin{figure*}
	  	\centering
	  	\includegraphics[width=0.78\linewidth]{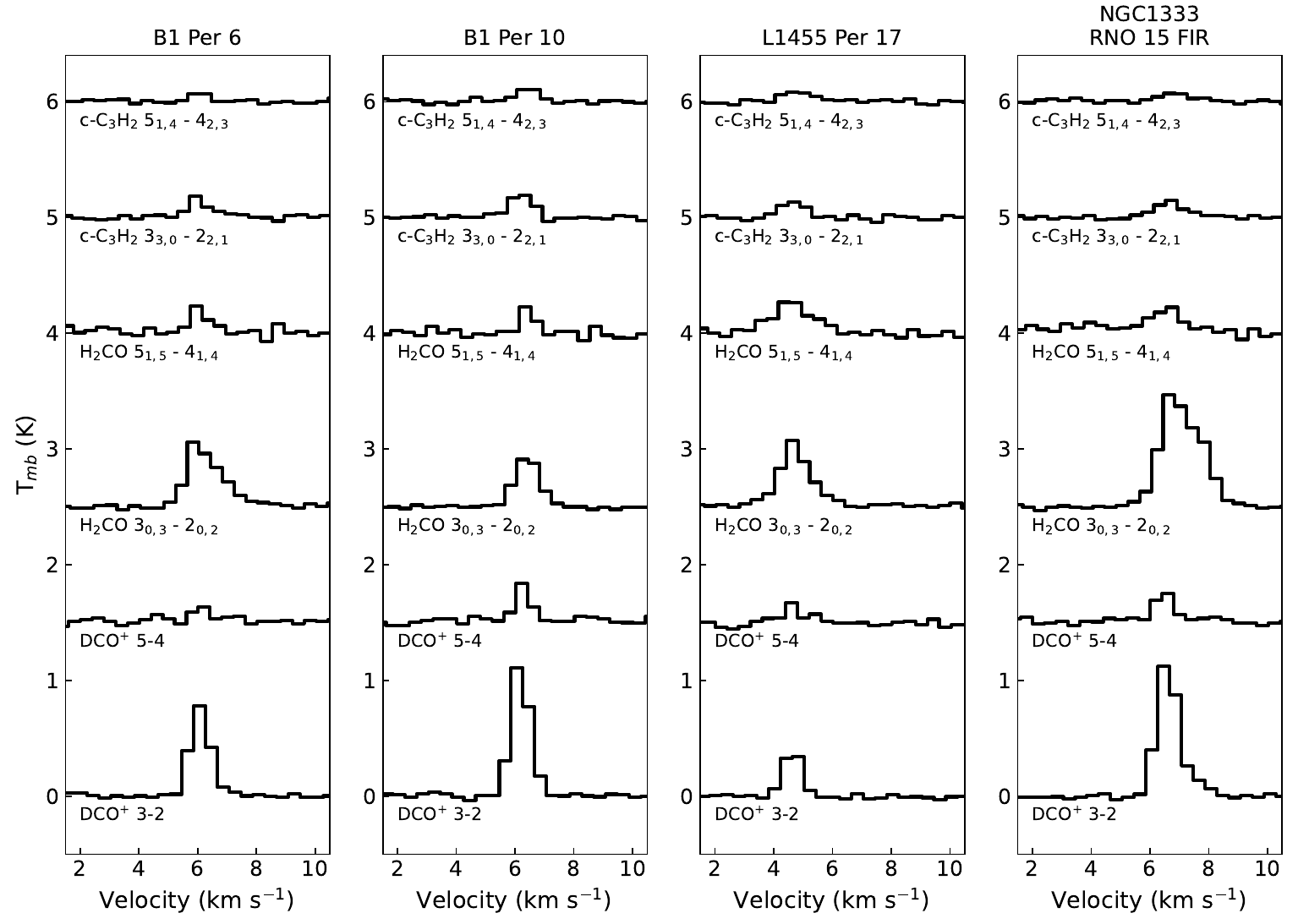}
	  	\caption{A sample of APEX spectra toward sources from the extended sample (see Sect.~\ref{sec:observations}). The spectra for \ce{DCO+} $J$=5--4 and \ce{p-H2CO} $J$=5$_{0,5}$--4$_{0,4}$ were obtained with APEX-2 with a beam of 18$\arcsec$, the other spectra were obtained with APEX-1 and a beam of 28.7$\arcsec$. From left to right, the first two panels show spectra for the components of the system B1 Per6+Per10. The third panel shows spectra for the close binary L1455 Per17. The fourth panel shows spectra for the single protostellar source NGC1333 RNO 15 FIR. The spectra are averaged to 0.4 km~s$^{-1}$ in order to increase sensitivity. All emission lines shown are found to have S/N $\geq$5 based on GILDAS/CLASS gaussian fitting routines.}
	  	\label{fig:APEX_spec}
	  \end{figure*}

\begin{figure*} 
	  	\centering
	  	\includegraphics[width=0.6\linewidth]{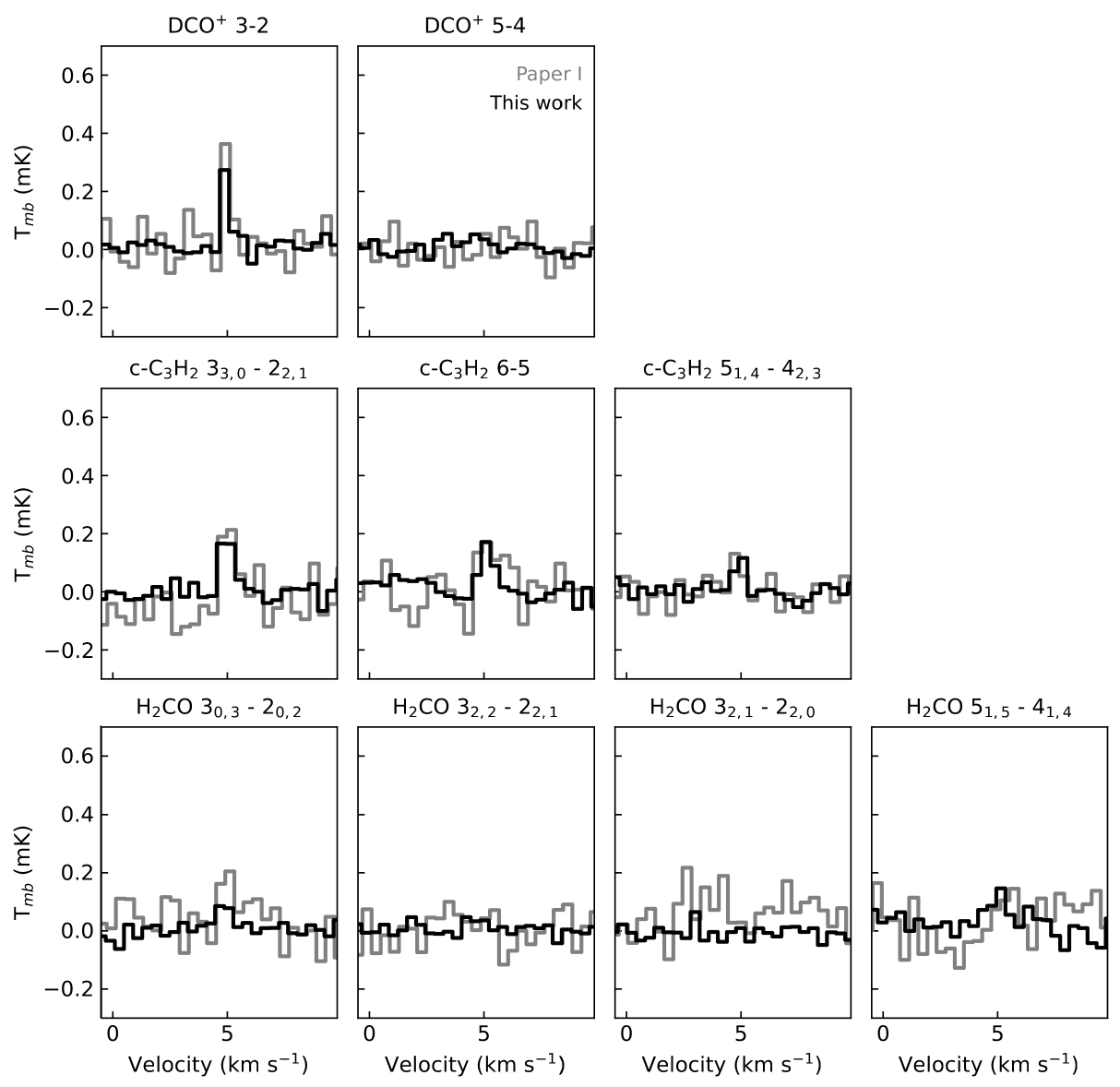}
	  	\caption{Comparison of spectra from the pilot (gray) and extended (black) samples for L1455 Per25. Both datasets were processed in the same manner and averaged to 0.4 km~s$^{-1}$. The molecular species shown are the same as listed in Table~\ref{tab:lines}. The extended sample observations provide a better constraint for single protostars than the pilot observations.}
	  	\label{fig:Per25_spec}
	  \end{figure*}

	  Samples of the Atacama Pathfinder EXperiment (APEX) spectra for the extended sample (Project codes O-0104.F-9307A and O-0104.F-9307B, see Sect.~\ref{subsec:APEXobs} for additional details) used in this work are presented in this appendix. Figure~\ref{fig:APEX_spec} shows spectra from three systems: a wide binary B1 Per6+Per10, a close binary L1455 Per17, and a single protostar NGC1333 RNO 15 FIR. Three molecular species with two transitions each are shown and were used to derive the physical parameters reported in this work: gas kinetic temperature from line peak ratios (see Sect.~\ref{subsec:Tgas}), column density (see Sect.~\ref{subsec:colden}), and gas masses (see Sect.~\ref{subsec:Mgas}).
	  The comparison spectra of L1455 Per25 obtained from the observations of the pilot \citep{murillo2018PaperI} and extended (this work) samples are shown in Fig.~\ref{fig:Per25_spec}. See Sect.~\ref{subsec:APEXobs} for more details.
	  
	\end{appendix}

\end{document}